\documentclass[%
 reprint,
superscriptaddress,
 amsmath,amssymb,
 aps,
 nofootinbib,
floatfix,
]{revtex4-1}

\usepackage{bm}
\usepackage{amsmath,amssymb}
\usepackage{amsfonts}
\usepackage{layouts}
\usepackage{mathtools}
\usepackage{dsfont}
\usepackage{graphicx}
\usepackage{float}
\usepackage{subfigure} 
\usepackage{verbatim}
\usepackage{amsfonts}
\usepackage{bbold}
\usepackage{dcolumn}
\usepackage{bm}
\usepackage{color}
\usepackage[dvipsnames]{xcolor}
\usepackage{listings}
\definecolor{blueprl}{RGB}{46,48,146}

\usepackage{mathrsfs}
\usepackage{filecontents}
\usepackage{times} 

\newcommand{\ket}[1]{\mbox{$| #1 \rangle$}}

\newcommand{\xdownarrow}[1]{%
  {\left\downarrow\vbox to #1{}\right.\kern-\nulldelimiterspace}
}

\makeatletter
\newcommand*{\balancecolsandclearpage}{%
  \close@column@grid
  \clearpage
  \twocolumngrid
}
\makeatother

\usepackage{gensymb}
\usepackage[breaklinks=true]{hyperref}
\hypersetup{
colorlinks   = true, 
urlcolor     = blue, 
linkcolor    = blue, 
citecolor    = blue 
}
\usepackage{comment}
\usepackage[normalem]{ulem}
\usepackage{xcolor}
\usepackage{graphicx}
\usepackage{cleveref}
\crefname{equation}{Eq.}{Eqs.}
\Crefname{equation}{Equation}{Equations}
\crefname{figure}{Fig.}{Figs.}
\Crefname{figure}{Figure}{Figures}
\crefname{figure}{Fig.}{Figs.}
\Crefname{figure}{Figure}{Figures}
\crefname{section}{Sec.}{Secs.}
\Crefname{section}{Section}{Sections}
\crefname{appendix}{Appendix}{Appendices}
\Crefname{appendix}{Appendix}{Appendices}
\crefname{table}{Table}{Tables}
\Crefname{table}{Table}{Tables}

\newcommand{\Tr}{\text{Tr}}
\newcommand{\tr}{\text{tr}}

\makeatletter
\def\@bibdataout@aps{%
 \immediate\write\@bibdataout{%
  @CONTROL{%
   apsrev41Control,author="08",editor="1",pages="0",title="0",year="1",eprint="1"%
  }%
 }%
 \if@filesw
  \immediate\write\@auxout{\string\citation{apsrev41Control}}%
 \fi
}%
\makeatother 

\begin{document}

\title{
Overcoming the repeaterless bound in continuous-variable quantum communication without quantum memories} 

\author{Matthew S. Winnel}\email{matthew.winnel@uqconnect.edu.au}
\affiliation{Centre for Quantum Computation and Communication Technology, School of Mathematics and Physics, University of Queensland, St Lucia, Queensland 4072, Australia}
\author{Joshua J. Guanzon} 
\affiliation{Centre for Quantum Computation and Communication Technology, School of Mathematics and Physics, University of Queensland, St Lucia, Queensland 4072, Australia}
\author{Nedasadat Hosseinidehaj} 
\affiliation{Centre for Quantum Computation and Communication Technology, School of Mathematics and Physics, University of Queensland, St Lucia, Queensland 4072, Australia}
\author{Timothy C. Ralph}
\affiliation{Centre for Quantum Computation and Communication Technology, School of Mathematics and Physics, University of Queensland, St Lucia, Queensland 4072, Australia}

\date{\today}

\begin{abstract}
One of the main problems in quantum communications is how to achieve high rates at long distances. Quantum repeaters, i.e., untrusted, intermediate relay stations, are necessary to overcome the repeaterless bound which sets the fundamental rate-distance limit of repeaterless communications. In this work, we introduce a continuous-variable protocol which overcomes the repeaterless bound and scales like the single-repeater bound using just one linear-optical device called a ``quantum scissor'', combining the entanglement distillation and entanglement swapping elements of previous repeater proposals into a single step, thus, removing the need for quantum memories. Implementing a standard continuous-variable quantum key distribution protocol using our repeater we predict key rates which surpass the repeaterless bound. Our protocol works well for non-ideal single-photon sources and non-ideal single-photon detectors, and can tolerate some level of excess noise, making our protocol implementable with existing technology. We show that our scheme can be extended to longer repeater chains using quantum memories, using less physical resources than previous schemes. Furthermore, for applications beyond key distribution, our scheme generalises to higher order and distils more entanglement at the cost of a reduced probability of success.
\end{abstract}

\maketitle


\section{Introduction}

Quantum communication~\cite{Gisin_2007} is the art of transferring quantum states from one place to another. A prominent application is quantum key distribution (QKD)~\cite{Pirandola_2020,Xu_2020} which is the task of sharing a secret random key between two distant parties. Whilst promising to solve the age-old problem of absolutely-secret communication~\cite{Mosca}, the rate at which secret key can be distributed is fundamentally limited by the transmission distance. Other cryptographic, computational and metrology applications of quantum communication are similarly limited.

Quantum repeaters~\cite{7010905,Muralidharan2016} promise to improve the performance of quantum communication tasks by dividing the total distance into shorter sections where photon loss and other noise can be managed more easily. The repeaterless bound~\cite{Pirandola_2017} sets the fundamental rate-distance limit and cannot be surpassed without a quantum repeater, also known as the Pirandola–Laurenza–Ottaviani–Banchi (PLOB) bound. This is generalised to include quantum-repeater chains~\cite{Pirandola_2019}, for instance, the fundamental single-repeater bound gives the rate-distance limit for protocols using one untrusted repeater station.

``Twin-field'' (TF) QKD~\cite{Lucamarini_2018} can overcome the PLOB bound and scales proportional to the single-repeater bound without complex repeater components such as quantum memories. TF QKD is based on discrete-variable (DV) systems but it deviates significantly from standard DV-QKD protocols, and other applications for it have not been identified. In this paper, we are concerned with continuous-variable (CV) systems~\cite{Weedbrook_2012} where the quantum information is encoded in an infinite-dimensional Hilbert space which is advantageous for QKD because Alice and Bob can use coherent states and efficient homodyne detection~\cite{Weedbrook_2004}. No simple CV protocol has been proposed that can beat the PLOB bound. It is important to find one because the ideal performance of CV QKD is better than the ideal performance of DV QKD since the Hilbert space is larger~\cite{Pirandola_2020}, and hopefully one can then approach the single-repeater bound.

Recent proposals for CV quantum repeaters~\cite{PhysRevA.95.022312,PhysRevA.98.032335,PhysRevResearch.2.013310,Dias_2020,Ghalaii_2020} are based on three essential ingredients: entanglement distribution, entanglement distillation, and entanglement swapping. First, entangled states are distributed between neighbouring nodes. Second, entanglement is distilled non-deterministically which overcomes loss and noise in the links. Quantum memories are required to hold onto the quantum states while neighbouring links succeed in distilling their entanglement. Third, joint measurements are performed on some of the modes, heralding entanglement between more-distant stations. Additional rounds of distillation and swapping can entangle stations separated over greater distances. This approach requires quantum memories to overcome the PLOB bound.

In this paper, we introduce a CV-repeater protocol which surpasses the PLOB bound with a simple architecture and without quantum memories. This is possible by combining the entanglement distillation and entanglement swapping elements of previous CV proposals into a single step using just one linear-optical device called a quantum scissor~\cite{pegg1998optical,ralph2009nondeterministic,xiang2010heralded,barbieri2011nondeterministic,PhysRevA.102.063715}. Scissors non-deterministically perform noiseless linear amplification (NLA) whilst truncating all higher-order Fock numbers (hence the name scissor). Scissors have proved useful in many quantum-communication schemes, for instance, to enhance point-to-point QKD~\cite{Ghalaii_2020_scissors} and for quantum repeaters~\cite{PhysRevResearch.2.013310,Dias_2020}. Previously, scissors have been used for the entanglement distillation step only, however, our protocol highlights that quantum scissors can in fact operate as loss-tolerant quantum relays by themselves. That is, they simultaneously perform entanglement distillation and entanglement swapping and the PLOB bound can be beaten without quantum memories.  Furthermore, our protocol can be extended to higher order and distil large amounts of entanglement at the cost of a reduced probability of success.

We demonstrate how our CV-repeater protocol can be used for CV QKD and compute asymptotic secret key rates secure against collective attacks for the simple no-switching protocol based on coherent states and heterodyne detection~\cite{Weedbrook_2004}. We give the eavesdropper (Eve) full control of the single-photon sources and single-photon detectors. Our protocol fixes the ``directional problem'' in CV QKD (as will be discussed later) which plagues, for example, the CV measurement-device-independent (MDI) protocol~\cite{Ma_2014,Zhang_2014,Ottaviani_2015,Pirandola2015}.  Using quantum memories, we show how our protocol can be extended into longer repeater chains which scale like the corresponding repeater bounds.


\section{Our CV-repeater protocol}

\begin{figure*}
\centering
\includegraphics[scale=1.2]{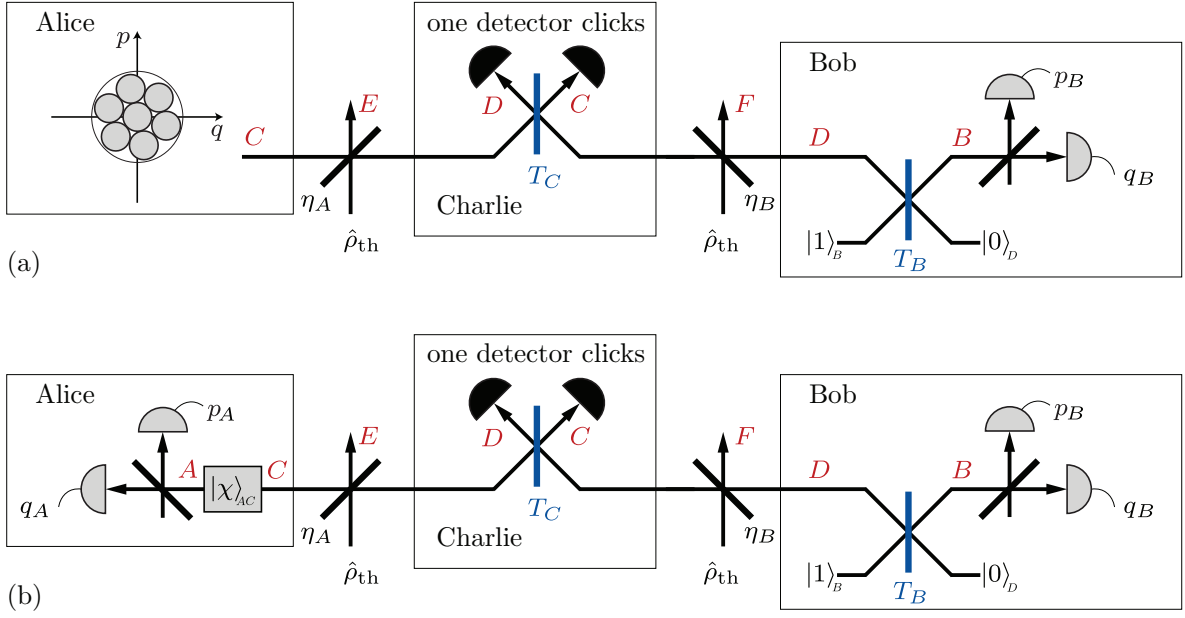}
\caption{Equivalent representations of our CV-QKD protocol for overcoming the PLOB bound. (a) Prepare-and-measure version. Alice chooses a coherent state at random from a two-dimensional Gaussian distribution and forwards it to Charlie down a thermal-noise channel. Meanwhile, Bob prepares a single-photon entangled state and sends one mode also towards Charlie. Charlie interferes the modes he receives and performs photon-number-resolving detection. A single click heralds strong correlations between Alice and Bob. Bob measures his remaining mode with heterodyne detection. The probability of success scales like the square root of the transmissivity of the total distance, thus, the repeater is effective at improving the rate-distance scaling. The gain of the NLA is tuned both by Charlie's exact location between Alice and Bob and the transmissivities of the beamsplitters, $T_C$ and $T_B$. (b) Entanglement-based (EB) version. Alice prepares an EPR state and performs heterodyne detection. The rest of the protocol is the same as the PM version.}
\label{fig:protocol}
\end{figure*}

We first consider our CV-repeater protocol for CV QKD, as shown in~\cref{fig:protocol}. The standard one-way Gaussian CV-QKD protocols are based on a Gaussian modulation of squeezed states or coherent states, and homodyne or heterodyne detection. We focus here on the coherent-state protocol without switching~\cite{Weedbrook_2004} since it is the simplest to implement. 

In the prepare-and-measure (PM) version, shown in~\cref{fig:protocol}(a), Alice prepares coherent states selected at random from a Gaussian modulation of variance $V_A$ to send to Bob. The channel of total transmissivity $\eta$ is divided into two shorter links each with transmissivity $\eta_A$ and $\eta_B$, such that $\eta=\eta_A \eta_B$, with an untrusted intermediate station, Charlie, between the trusted parties, Alice and Bob. Bob's prepared entangled state and Charlie's station together form a single-photon quantum scissor~\cite{pegg1998optical,ralph2009nondeterministic,barbieri2011nondeterministic}. When Charlie registers just a single click at one of his detectors, strong correlations are heralded between Alice and Bob resulting in a virtually lossless communication channel. Bob performs heterodyne measurements on his final states. His data are correlated with the states Alice sent and after classical post-processing and privacy amplification they share a secret random key.

The entanglement-based (EB) version is shown in~\cref{fig:protocol}(b) which is used for security analysis of the PM version or for entanglement distillation. Alice initially prepares an EPR state (two-mode-squeezed-state), $\ket{\chi}_{AC} {=}  \sqrt{1{-}\chi^2} \sum_{n=0}^\infty \chi^n \ket{n}_A\ket{n}_C$, with two-mode-squeezing parameter $r$, variance $\nu{=}\cosh{2r}$, $\chi{=}\tanh{r}$, and mean photon number $\bar{n}{=}\sinh^2{r}$, and where $\ket{n}$ are Fock-number states. The modulation variance $V_A$ in the PM version is related to $\chi$ in the EB version by $\nu{=}V_A{+}1{=}\cosh{(2\tanh^{-1}{\chi})}$. The two versions are equivalent if Alice performs heterodyne detection on her mode.

Quantum scissors~\cite{PhysRevA.102.063715} perform noiseless linear amplification (and deamplification) on arbitrary input states up to some Fock number $k$. They work by combining the input state with an entanglement-resource state and post selecting on $k$ detectors registering a single click, teleporting and amplifying the input state onto the outgoing mode.  The transmissivity of Bob's beamsplitter, $T_B$, sets the gain. Let's now consider loss between Bob and Charlie. Remarkably, we note here that quantum scissors perform linear amplification sufficiently well despite the loss if the input states are selected from a classical mixture but now with a modified gain which depends on the loss. We say that the scissor is loss tolerant. This result was mentioned previously with reference to more general scissor-like devices called ``tele-amplification''~\cite{neergaard-nielsen2013quantum}, and these devices are loss tolerant if the input states are restricted to a classical mixture of coherent states on a ring in phase space. A classical mixture is precisely what Alice sends towards Bob for Gaussian-modulated CV QKD, thus, quantum scissors are suitable as repeaters for CV QKD. We call it a repeater since it actually ``repeats'' in the sense that it beats the PLOB bound.


When placed halfway between Alice and Bob, the success probability of the repeater scales as the square root of the total channel transmissivity so the protocol can surpass the PLOB bound which scales with total channel transmissivity. This is in contrast to CV-MDI QKD~\cite{Pirandola2015} which cannot beat the PLOB bound (the MDI relay is not effective at ``repeating''). The secret key rate for an ideal implementation of the coherent-state protocol using our CV repeater is plotted in~\cref{fig:rate} and we beat the bound at 223 km.

We introduced the repeater in terms of teleportation and NLA, however, it can also be viewed as hybrid entanglement swapping where DV entanglement is used to fix the ``directional problem'' in CV QKD. The directional problem is that reverse reconciliation works much better than direct reconciliation because the reference should be the more noisy state (i.e., Bob). This means that the CV-MDI-QKD protocol~\cite{Pirandola2015} works extremely asymmetrically. It also means that recent repeater proposals~\cite{PhysRevResearch.2.013310,Dias_2020,Ghalaii_2020} are set-up such that they always work better if the states are propagated down lossy channels in a direction from Alice towards Bob and not in the other direction.

The gain of the NLA is tuned by changing the exact location of the repeater between Alice and Bob (i.e., varying $\eta_A$ and $\eta_B$) as well as in the traditional way~\cite{ralph2009nondeterministic} of tuning the transmissivities of the beamsplitters in the scissor, $T_C$ and $T_B$. We in general let Charlie's beampslitter be balanced, $T_C=1/2$, then the gain of the repeater for pure loss is $g = \sqrt{\frac{ T_B}{\eta_B(1-T_B)}}$. This is the same gain relation for loss-tolerant tele-amplification, see Eq. 5 of Ref.~\cite{neergaard-nielsen2013quantum}. Putting this all together, in our CV-repeater protocol, Alice's state experiences loss which is amplified by the scissor where the gain $g$ is tuned by the location of Charlie and the transmissivity of Bob's beamsplitter.

For pure loss, the global output state shared between Alice, Bob, and the environment, heralded by a single click at the mode-$D$ detector, is
\begin{multline}
|\psi\rangle_{ABEF} = \sqrt{\frac{1{-}\chi^2}{2}} \sum_{k=0}^\infty  (1{-}\eta_A)^\frac{k}{2} \\ \biggl[    \chi^k \sqrt{\eta_B}  \sqrt{1{-}T_B}  \ket{k}_A\ket{0}_B\ket{k}_E\ket{0}_F \\  + \chi^{k{+}1} \sqrt{k{+}1} \sqrt{\eta_A}  \bigl(\sqrt{T_B}  \ket{k{+}1}_A\ket{1}_B\ket{k}_E\ket{0}_F \\ + \sqrt{1{-}\eta_B} \sqrt{1{-}T_B} \ket{k{+}1}_A\ket{0}_B\ket{k}_E\ket{1}_F \bigr)  \biggr].\label{eq:output}
\end{multline}
$k$ is the number of photons lost in Alice's link. For each $k$, the scissor operates ideally and noiselessly for the first two terms, however, for the third term, mode $D$ lost one photon to mode $F$ and this is an error. Thus, there are three types of noise on the global output state: loss on Alice's link, truncation noise from the scissor, and decoherence due to the error term when the scissor loses a photon.

Alice and Bob's final state $\hat{\rho}_{AB}$ is obtained by tracing over the environmental modes, $E$ and $F$. The total success probability is $P = 2\tr({\hat{\rho}_{AB}})$, where the factor of two is because there are two successful click patterns, one of which heralds a passive $\pi$-phase shift on the output state and is easily corrected~\cite{PhysRevA.102.063715}, Charlie simply tells Bob which detector fired.  We present the derivation of~\cref{eq:output} in~\cref{sec:protocol}, where we also consider thermal noise and experimental imperfections.

\section{Computation of the secret key rate}

The asymptotic  secret key rate $K$ is given by the raw key, $K_\text{raw}$, multiplied by the rate  of successful operation of the repeater protocol, $R$. That is, $K= R K_\text{raw}$. The raw key is given by the asymptotic secret key rate formula, $K_\text{raw}{=}\beta I_{AB}{-}\chi_{EB}$~\cite{devetak2005distillation}, where $I_{AB}$ is Alice and Bob's classical mutual information, $\beta$ is the reconciliation efficiency, and $\chi_{EB}$ is an upper bound on Eve's maximal information.

Eve's information is upper bounded by $\chi_{EB}$ which can be bounded as a function of the heralded covariance matrix, $\Gamma_{AB}$, in the EB version of the protocol. Since Alice's modulation is Gaussian, $\Gamma_{AB}$ is directly accessible by Alice and Bob if they perform the PM version of the experiment. From Alice and Bob's total data in the PM scheme, they can reconstruct the equivalent EB scheme. Then they use their reconstructed EB version to estimate $\Gamma_{AB}$, and use Gaussian optimality~\cite{PhysRevLett.96.080502} to upper-bound Eve's information $\chi_{EB}$. No assumption on the channel is required in an experiment for asymptotic security. We refer you to~\cref{sec:security} for more details.

However, since we do not have access to experimental data, we assume ambient conditions to simulate the parameters accessible to Alice and Bob in an experiment. We consider that the so-called ``excess noise'' is coming from the environment as input thermal noise ($\hat{\rho}_{\text{th}}$) with variance $V$. Consider Ref.~\cite{PhysRevLett.125.010502} which performed a long-distance CV-QKD experiment. They had an input excess noise $\xi=0.0081$ shot noise units (SNU) for 32.45 dB of loss (i.e., 162.25 km at 0.2dB/km). The distances involved in our system are much greater than 100 km, so we consider an amount of input thermal noise such that an equivalent amount of excess noise on the direct transmission system is $\xi=0.02$ SNU at 350 km (70 dB). This noise with variance $V$ we inject evenly into all links. That is, the input thermal noise variance is fixed and excess noise $\xi$ builds up over long distances. All key rates plotted in this paper have this amount of thermal noise.

\begin{figure}
\includegraphics[width=1\linewidth]{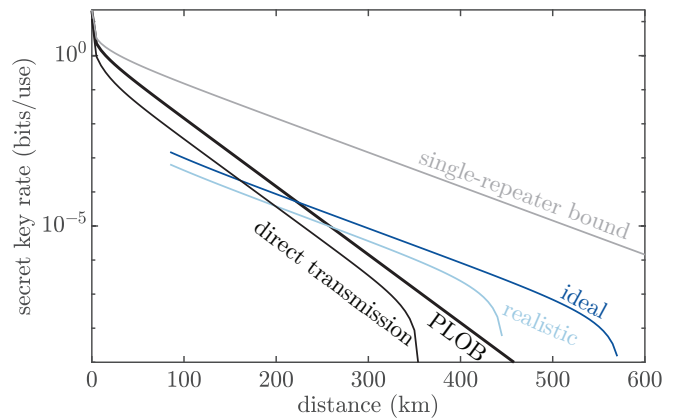}
\caption{Secret key rate of our CV-QKD protocol based on coherent states and heterodyne detection versus the total distance between Alice and Bob assuming standard optical fibre (0.2 dB/km) and excess noise. The amount of excess noise is defined in the text. The strength of Alice's EPR state is $\chi=0.4$ ($V_A=0.38$ in the PM version), and the beamsplitters and location of the relay is roughly optimised. The reconciliation efficiency is $\beta= 0.95$. For the realistic implementation (light blue), we assume $75\%$ single-photon source and single-photon detector efficiencies and $10^{-8}$ dark-count rate probability, whilst these are assumed perfect for ideal (dark blue). We plot direct transmission for optimised modulation variance and with excess noise. We also plot the point-to-point PLOB bound~\cite{Pirandola_2017} and the single-repeater bound~\cite{Pirandola_2019}. Also note that the squeezed-state protocol and homodyne detection can increase the key rate, achieving similar rates as the best variants of TF QKD.}
\label{fig:rate}
\end{figure}

The ideal secret key rate of our protocol based on coherent states and heterodyne detection is shown in~\cref{fig:rate} for fixed thermal noise (the amount defined above). The modulation variance is $\chi=0.4$ and the relay is placed off centre, slightly closer to Bob than Alice, which gives good key rates and is probably close to optimal (we cannot numerically compute key rates for arbitrarily large $\chi$). The reconciliation efficiency is $\beta=0.95$. For comparison, we plot the key rate for direct transmission for optimised modulation variance. The PLOB~\cite{Pirandola_2017} and single-repeater~\cite{Pirandola_2019} bounds are also shown. These bounds are $-\log_2{(1-\eta^{1/N})}$ where $N$ is the number of links dividing the total distance, i.e., the number of repeaters is $N{-}1$ (for the PLOB bound $N{=}1$ and for the single-repeater bound $N{=}2$). Our ideal protocol beats the PLOB bound at 223 km and scales like the single-repeater bound. It beats direct transmission at 166 km.

Scissors are robust to non-ideal single-photon sources and detectors (and even on-off detectors) in the high fidelity regime of operation, i.e., low-energy input states~\cite{PhysRevA.102.063715}. These experimental imperfections generally decrease the success probability but do not greatly affect the fidelity. We plot the secret key rate for a realistic implementation in~\cref{fig:rate} assuming $75\%$ single-photon source and single-photon detector efficiencies and $10^{-8}$ dark-count rate probability. The realistic curve beats the PLOB bound at 260 km and direct transmission at 203 km.

Placing the repeater symmetrically between Alice and Bob, $\eta_A=\eta_B=\sqrt{\eta}$ and $T_C=T_B=0.5$, also gives good key rates. We refer to this set-up as the symmetric configuration (here, $\chi = 0.1$ is about optimal). However, we find that by moving the repeater slightly closer to Bob and increasing the variance of the state prepared by Alice, and tuning the beamsplitters accordingly, the success probability is increased (at the expense of some decrease of the raw key), thus, improving the secret key rate overall. It is the asymmetric configuration which we plot in the figure noting that the symmetric configuration would decrease the key rate such that the PLOB bound is beaten at about 50 km greater distance. Also, CV QKD based on squeezed states and homodyne detection~\cite{PhysRevA.63.052311} can increase the key a fair amount, obtaining similar rates achievable by the best TF-QKD variants.

\section{Entanglement distillation with higher-order scissors}

Quantum scissors generalise to higher order at the cost of a reduced success probability~\cite{PhysRevA.102.063715}. For some applications, where entanglement distillation is more important than rate, for instance, for teleportation protocols and distributed quantum computing, higher-order quantum scissors may prove useful. When used as a repeater, like in~\cref{fig:protocol}, the three-photon scissor cannot beat the PLOB bound, however, it distills a large amount of entanglement, more than the single-photon scissor can achieve. The single-photon repeater scales like $ \sqrt{\eta}/2$ whilst the three-photon repeater scales like $\frac{3}{64} {\eta}^\frac{3}{2}$. The prefactor is a penalty due to using linear optics, and waiting for the required number of clicks in the detector.

\begin{figure}
\includegraphics[width=1\linewidth]{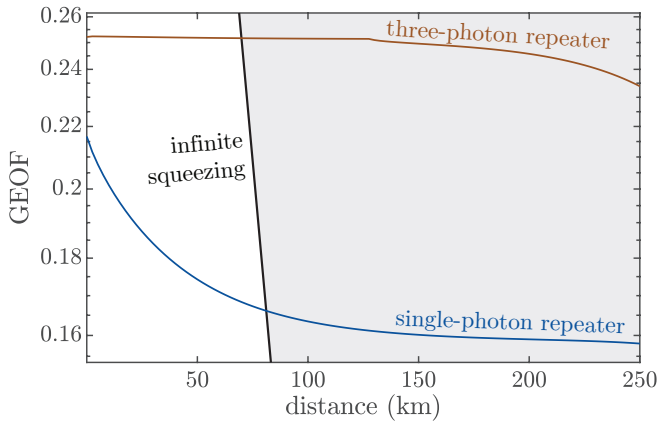}
\caption{Gaussian entanglement of formation (GEOF) versus distance for the repeater protocol shown in~\cref{fig:protocol}(b) in symmetric configuration ($T_C{=}T_B{=}0.5,\eta_A{=}\eta_B{=}\sqrt{\eta}$), $\chi=0.2$, and with excess noise. We also plot the GEOF for an EPR state with infinite squeezing. The shaded region is unattainable by any deterministic protocol. It is clear from this figure that higher-order repeaters distil more entanglement and work for larger-energy input states, beyond what is possible for the first-order repeater.}
\label{fig:GEOF_3scissor}
\end{figure}

As an entanglement measure, we calculate the Gaussian entanglement of formation (GEOF)~\cite{PhysRevA.69.052320,PhysRevA.96.062338} of the final state, $\hat{\rho}_{AB}$, to evaluate the performance of the first- and third-order repeater in symmetric configuration.  We plot this in~\cref{fig:GEOF_3scissor} for $\chi=0.2$. We use the GEOF since for a given covariance matrix the entanglement is minimised by Gaussian states meaning that we will not overestimate the amount of entanglement. For small $\chi$, the repeater introduces only a little non-Gaussian noise. We note in the figure that the entanglement is barely unchanged with distance.

We also note that in the high-fidelity regime the output state remains very pure as a function of distance, whereas, the infinitely-squeezed EPR state exponentially loses its purity with distance. For instance, the purity of the output state heralded by the relay in symmetric configuration is greater than about $0.95$ for $\chi{=}0.1$ and $0.85$ for $\chi{=}0.2$. We refer you to~\cref{sec:higherscissors} for more details.

\section{Repeater chain using quantum memories}

\begin{figure*}
\includegraphics[width=1\linewidth]{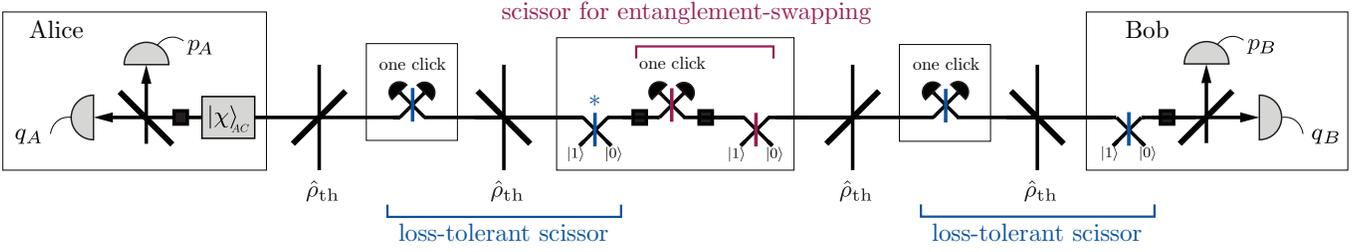}
\caption{Extending our CV-QKD repeater protocol into a three-repeater chain using quantum memories (solid squares) and additional quantum scissors. Shown here is the entanglement-based version. At the lowest level, the quantum scissors (shown in blue) perform entanglement swapping and entanglement distillation (without the need for quantum memories) and are tolerant to loss in the links. At the higher level, quantum memories hold onto the state of the lower level and an additional quantum scissor (shown in red) (i.e., a partial Bell measurement) performs entanglement swapping. The secret key rate scales of order $\eta^{1/4}$, the same scaling as the fundamental three-repeater bound. This repeater protocol requires far less resources than previous CV schemes~\cite{PhysRevResearch.2.013310,Dias_2020,Ghalaii_2020}. Note that quantum scissors have two successful click patterns, one of which heralds a $\pi$-phase shift on the output mode~\cite{PhysRevA.102.063715}. This phase shift is easily corrected via passive phase shifts but they are not shown in the figure for simplicity.}
\label{fig:repeater}
\end{figure*}

\begin{figure}
\includegraphics[width=1\linewidth]{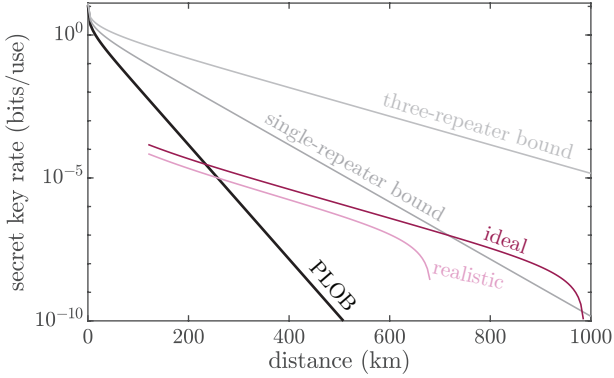}
\caption{Secret key rate versus the total distance between Alice and Bob with excess noise of our three-repeater chain for the no-switching protocol based on coherent states and heterodyne detection. The repeater chain is set-up in a slightly asymmetric configuration with $\chi=0.2$ as to roughly optimise the key rate. The realistic implementation (pink) assumes $75\%$ single-photon source and single-photon detector efficiencies and $10^{-8}$ dark-count rate probability. The quantum memories are assumed to be ideal.}
\label{fig:rate_4links}
\end{figure}

We extend our single-node protocol to a three-repeater chain in~\cref{fig:repeater} and we plot the secret key rate in~\cref{fig:rate_4links}. It scales like the three-repeater bound, ${-}\log_2{(1{-}\eta^{1/4})}$~\cite{Pirandola_2019}.  We achieve the same scaling with half the number of resources as previous CV-repeater proposals~\cite{PhysRevResearch.2.013310,Dias_2020,Ghalaii_2020}. Our ideal three-repeater chain beats the PLOB bound at 234 km and the single-repeater bound at 717 km.

For the realistic implementation, we assume $75\%$ single-photon source and single-photon detector efficiencies and $10^{-8}$ dark-count rate probability. The quantum memories are assumed ideal. Details for computing secret key rates of our three-repeater chain are presented in~\cref{sec:three-repeater}. The chain can straightforwardly be extended to include more repeaters in a longer repeater chain.

\section{Discussion and Conclusion}
In this paper, we introduced a practical CV-quantum communication protocol based on a quantum scissor which when used for CV QKD overcomes the PLOB bound without the need for quantum memories. The scissor is used as a loss-tolerant quantum repeater. It is useful for QKD and for entanglement distillation of EPR states. Equivalently, the scheme can be thought of as hybrid entanglement swapping, using CV and DV entanglement. The DV entanglement fixes the directional problem in CV QKD. The single-photon measurement is a powerful non-Gaussian resource, projecting onto a state with strong correlations between Alice and Bob allowing the generation of a secret key or the distillation of entanglement.

The position of the repeater and the transmissivities of the beamsplitters together set the gain of the NLA. By placing the repeater symmetrically between Alice and Bob and using balanced beamsplitters, the success probability scales of order $\sqrt{\eta}$ and the gain of the NLA across the second link overcomes the loss in the first link. Thus, the state is sufficiently teleported and amplified from Charlie to Bob at a $\sqrt{\eta}$-scaling probability of success.

The protocol is tolerant to some excess noise and imperfections of the single-photon sources and single-photon detectors. The protocol can be extended into a chain of quantum repeaters using quantum memories. This would require just half the number of resources as recent CV quantum-repeater architectures~\cite{PhysRevResearch.2.013310,Dias_2020,Ghalaii_2020}.

One of the main results of this paper is that we in general improve the scaling of many recent protocols based on quantum scissors to the square root of their old scaling and, actually, our protocol is quite robust to excess noise. This promotes the success probability of higher-order scissors into regimes of practical use.

It is also interesting to note that TF QKD requires the two links to have similar levels of loss~\cite{Zhong2021}, whereas, our protocol works both symmetrically and asymmetrically by optimising the gain (but not beating the PLOB bound in the very asymmetric case). One advantage of TF QKD is that the only measurement is performed in Charlie's station and so can be controlled by Eve. For our protocol, we have a CV measurement in Bob's station which we assume is trusted.

An open question is how to approach the single-repeater bound without quantum memories, that is, how to increase the size of the modulation without introducing more truncation noise from the scissor since approaching the single-repeater bound requires very large modulations. An interesting feature would be to generalise our protocol to reliably connect multiple trusted users. Future work would involve extending security analysis to the finite-size regime.


\begin{acknowledgements}

We thank Josephine Dias for valuable comments during our investigation. This research was supported by the Australian Research Council (ARC) under the Centre of Excellence for Quantum Computation and Communication Technology.

\end{acknowledgements}

\onecolumngrid
\appendix


\section{\label{sec:protocol}Our CV-repeater protocol for overcoming the repeaterless bound}

In this section, we describe how to calculate the final output state of the entanglement-based (EB) version of our single-repeater CV-QKD protocol for overcoming the repeaterless bound, shown in~\cref{fig:protocol}(b) of the main text.

\subsection{Ideal implementation for pure loss}

We first consider the ideal implementation of our protocol, consisting of a perfect single-photon source and perfect single-photon detectors. Initially, Alice and Bob each prepare an entangled resource state. Alice prepares an EPR state $\ket{\chi}_{AC}$, while Bob prepares a single photon mixed with vacuum on a beamsplitter with transmissivity $T_B$. The beamsplitter transformation is
\begin{align}
    \hat{B}(\theta) &= e^{i{\theta}(\hat{a}^\dagger \hat{b} + \hat{b} \hat{a}^\dagger)},
\end{align}
where $\hat{a}$ and $\hat{b}$ are the annihilation operators of the two modes, and the transmissivity is determined by $T_B = \cos^2{\theta}$. Hence, Alice and Bob's initial state is
\begin{align}
    |\psi_0\rangle_{ACDB} &= \ket{\chi}_{AC} ( \sqrt{T_B} \ket{0}_D\ket{1}_B + \sqrt{1{-}T_B} \ket{1}_D\ket{0}_B ).
\end{align}

Modes $C$ and $D$ are individually propagated through different lossy channels with transmissivity $\eta_A$ and $\eta_B$, respectively. The lossy channels are modelled by introducing environmental modes $E$ and $F$, initially in the vacuum state (or a thermal state to model excess noise), and mixed on a beamsplitter. After the pure-lossy links, the state becomes
\begin{multline}
|\psi_0\rangle_{ACEDBF} \to  \sqrt{1{-}\chi^2} \sum_{n=0}^\infty \chi^n \sum_{k=0}^n \sqrt{{n \choose k}}
(1{-}\eta_A)^{\frac{k}{2}} \eta_A^{\frac{n-k}{2}}|n\rangle_{A} |n-k\rangle_{C} |k\rangle_E\\
 \biggl[ \sqrt{T_B} \ket{0}_D\ket{1}_B \ket{0}_F +  \sqrt{1{-}T_B} \bigl ( \sqrt{\eta_B} \ket{1}_D\ket{0}_B  \ket{0}_F + \sqrt{1{-}\eta_B}\ket{0}_D\ket{0}_B \ket{1}_F \bigl )\biggr].
\end{multline}

Modes $C$ and $D$ are combined on a beamsplitter, which we assume with no loss of generality a transmissivity $T_C=1/2$, and measured with single-photon detectors. We post-select on the instances where the mode-$D$ detector registers a single photon and the mode-$C$ detector registers vacuum (i.e. $\langle 0|_C \langle 1|_D$). The output state is
\begin{multline}
|\psi\rangle_{ABEF} = \sqrt{\frac{1{-}\chi^2}{2}} \sum_{k=0}^\infty  (1{-}\eta_A)^\frac{k}{2} \biggl[    \chi^k \sqrt{\eta_B}  \sqrt{1{-}T_B}  \ket{k}_A\ket{0}_B\ket{k}_E\ket{0}_F \\ + \chi^{k{+}1} \sqrt{k{+}1} \sqrt{\eta_A}  \bigl(\sqrt{T_B}  \ket{k{+}1}_A\ket{1}_B\ket{k}_E\ket{0}_F + \sqrt{1{-}\eta_B} \sqrt{1{-}T_B} \ket{k{+}1}_A\ket{0}_B\ket{k}_E\ket{1}_F \bigr)  \biggr].
\end{multline}

Alice and Bob's final state $\hat{\rho}_{AB}$ is obtained by tracing out the environmental modes, $E$ and $F$. The probability that this protocol succeeds is $P=2\tr({\hat{\rho}_{AB}})$ where the factor of $2$ accounts for the other click pattern (i.e. $\langle 1|_C \langle 0|_D$), which heralds a passive $\pi$-phase shift on the output state, easily correctable by Bob.

In the symmetric configuration, the repeater is placed exactly in the centre between Alice and Bob. In this case, we expect symmetric loss $\eta_A=\eta_B=\sqrt{\eta}$, such that it's best to set $T_C=T_B=1/2$, which results in a final output state 
\begin{multline}
|\psi (\text{symmetric})\rangle_{ABEF} = \sqrt{\frac{1{-}\chi^2}{2}} {\eta}^\frac{1}{4} \sum_{k=0}^\infty  (1{-}\sqrt{\eta})^\frac{k}{2} 
\biggl[    \chi^k    \ket{k}_A\ket{0}_B\ket{k}_E\ket{0}_F \\ +   \chi^{k{+}1} \sqrt{k{+}1}  \bigl( \ket{k{+}1}_A\ket{1}_B\ket{k}_E\ket{0}_F + \sqrt{1{-}\sqrt{\eta}} \ket{k{+}1}_A\ket{0}_B\ket{k}_E\ket{1}_F \bigr)  \biggr].
\end{multline}
It is important to consider how the success probability $P$ depends on the transmissivity of the channel $\eta$. In the limit of large distances (i.e. small $\eta$ transmissivity) and small $\chi$ input states, the success probability scales as $P \propto \sqrt{\eta}/2$. Thus, the secret key rate also scales proportional to $\sqrt{\eta}$, and hence the protocol can overcome the PLOB bound which scales like $\eta$.

Let us now consider the output state more carefully. The number of lost photons from mode $C$ to mode $E$ is given by $k$. There is also loss in the scissor itself, from mode $D$ to mode $F$. The first term corresponds to the instance where Bob's photon clicked the detector, while the second term is where Alice's photon clicked the detector. In particular, no photons were lost from mode $D$ for the first two terms, which represent the states in which the scissor operated perfectly. The third term is an error state, where mode $D$ lost one photon to mode $F$.

Thus, there are three types of errors: loss on Alice's link (mode $E$), truncation noise because of the quantum scissor, and decoherence because of the loss \textit{in} the scissor (mode $F$). The loss on Alice's link is overcome by increasing the gain $g$ of the scissor. The truncation noise is minimised by choosing the input variance $V_A$ and gain $g$ so that the output state is mostly a superposition of a single photon and vacuum. The decoherence term is unavoidable due to the loss in the scissor, however, as we stress in the main text, using Bob as the reference for reconciliation, we still get a positive key rate.

\subsection{Asymmetric configuration}

In~\cref{fig:rate} of the main text, we tune the beamsplitters and $\chi$ so that the key rate is roughly as optimal as possible. Specifically, the energy of Alice's EPR state is $\chi=0.4$ ($V_A=0.38$ in the PM version), the beamsplitter transmissivities in the scissor are $T_C=1/2$ and $T_B=2/3$, and the location of the repeater is moved slightly towards Bob such that $\sqrt{\eta_A}g=0.21$, where $g = \sqrt{\frac{ T_B}{\eta_B(1-T_B)}}$. This means the gain does not amplify all the way back up to the energy of the input state; this is necessary because the modulation here is large and the scissor introduces too much truncation noise for larger gain. This configuration is approximately optimal. There is a trade-off between success probability and entanglement distillation.

\subsection{Excess noise and experimental imperfections}

In this section, we detail how we incorporated excess noise and experimental imperfections in our protocol.

In CV QKD, noise on top of pure loss is called ``excess noise''. We model this additional noise by considering thermal-state inputs from the environment, instead of vacuum as in the pure loss case. The variance of the input thermal noise is chosen such that for direct transmission it is equivalent to ``excess noise'' $\xi= 0.02$ SNU at 70 dB loss (that is, 350 km at 0.2 dB/km). This amount of noise is probably what can be expected in future demonstrations, for instance, consider the long-distance CV-QKD experiment over 202.81 km from Ref.~\cite{PhysRevLett.125.010502}. At their longest distance, they have an excess noise of $\xi= 0.0081$ SNU at 32.45 dB loss (that is, 162.25 km at 0.2 dB/km).

We model the efficiency of the non-ideal single-photon source by placing a beamsplitter of transmissivity $\tau_\text{s}$ after the source (here, the transmissivity is the efficiency). We model the detection efficiency of the single-photon detectors by placing a beamsplitter of transmissivity $\tau_\text{d}$ before each detector. To model dark counts, we assume a thermal state of mean photon number $\bar{n}_d$ is incident on the auxiliary beamsplitter port and choose $\bar{n}_d$ such that the required dark-count rate is achieved. For each $\tau_\text{d}$ there is a $\bar{n}_d$ that gives a specific dark-count rate.

In order to compute secrete key rates for noisy channels and to include experimental imperfections, we do not analytically calculate the output density matrix $\hat{\rho}_{AB}$, rather, we do it numerically in Fock space.

\section{\label{sec:security}Computation of the secret key rate}

In this section, we describe how to compute asymptotic secret key rates and we discuss security analysis.

\subsection{Asymptotic secret key rate formula}

The secret key rate is given by
\begin{align}
    K =  R  K_\text{raw},
\end{align}
where $R$ is the rate of successful operation of the repeater protocol and $K_\text{raw}$ is the raw key. For our single-repeater protocol, $R$ is simply the success probability, $P$, of the repeater detecting a single click, $R=P=2\tr(\hat{\rho}_{AB})$. The raw key, $K_\text{raw}$, is given by the asymptotic secret key rate formula $K_\text{raw}=\beta I_{AB}-\chi_{EB}$~\cite{devetak2005distillation}, where $I_{AB}$ is Alice and Bob's classical mutual information, $\beta$ is the reconciliation efficiency, and $\chi_{EB}$ is an upper bound on Eve's maximal information (the Holevo bound with Bob who is the reference side of the information reconciliation).

In what follows, we show that Alice and Bob's mutual information can be lower bounded and Eve's maximal information can be upper bounded using only the covariance matrix shared between Alice and Bob in the EB version, thereby giving a lower bound on the secret key rate. Since the output state $\hat{\rho}_{AB}$ is close to Gaussian, the exact key rate should be close to our lower bound. The output state $\hat{\rho}_{AB}$, due to symmetry of the protocol, has covariance matrix of the following form:
\begin{align}
   \Gamma_{AB} &= \begin{bmatrix}
        a & 0 & c & 0\\
        0 & a & 0 & -c\\
        c & 0 & b & 0\\
        0 & -c & 0 & b\\
    \end{bmatrix}.\label{eq:CM}
\end{align}

\subsection{Alice and Bob's mutual information}

The mutual information quantifies the amount of correlations between Alice and Bob, measured in bits of correlation. It is given by~\cite{Sanchez2007QuantumIW}
\begin{align}
    I_{AB} &= H(A)-H(A|B),
\end{align}
where $H(\cdot)$ is the Shannon entropy. The notation $H({A|B})$ means the conditional Shannon entropy of Alice's data conditioned on Bob's measurements.

The output state is close to a Gaussian state so we can approximate Alice and Bob's information $I_{AB}$ using the covariance matrix $\Gamma_{AB}$ shared between Alice and Bob. For the protocol based on coherent states and heterodyne detection, we have 
\begin{align}
    I_{AB} &\approx  \log_2 \left({ \frac{V_{A}}{V_{A|B}}}\right) = \log_2 \left({ \frac{1+a}{1+a-\frac{c^2}{1+b}}}\right),\label{eq:I_AB}
\end{align}
where $a,b,$ and $c$ are elements of the covariance matrix in~\cref{eq:CM}.
Numerical calculations confirm that this actually underestimates Alice and Bob's information, hence the key rate is secure using $\Gamma_{AB}$ to approximate $I_{AB}$.

\subsection{Upper bounding Eve's information}

Eve's maximal information with Bob is given by the Holevo quantity~\cite{holevo1973bounds}
\begin{equation}
\chi_{EB} = S(\hat{\rho}_E) - S(\hat{\rho}_{E|b}),\label{eq:HOLEVO}
\end{equation}
where $S(\hat{\rho}_E)$ is the von Neumann entropy of Eve's state, and $S(\hat{\rho}_{E|b})$ is the von Neumann entropy of Eve's state conditioned on Bob's measurement. 

However, we cannot calculate $\chi_{EB}$ directly because we do not know Eve's optimal attack since the protocol is both non-Gaussian and non-deterministic. Since Alice's initial modulation is Gaussian, Alice and Bob have access to a covariance matrix $\Gamma_{AB}$ in the EB version from parameters they observe in the PM version. In order to compute secret key rates, we can simulate $\Gamma_{AB}$ assuming a thermal-lossy channel. Given $\Gamma_{AB}$, estimated by Alice and Bob in the simulated experiment, we can then bound Eve's information using Gaussian optimality~\cite{PhysRevLett.96.080502} which says that for an arbitrary quantum state $\hat{\rho}_{AB}$ shared between Alice and Bob, the Gaussian state $\hat{\rho}^*_{AB}$ with the same covariance matrix $\Gamma_{AB}$ as for $\hat{\rho}_{AB}$ gives the maximal Holevo information. Thus, Eve's information can be upper bounded given the covariance matrix $\Gamma_{AB}$ shared between Alice and Bob, even if the state is non-Gaussian. Explicitly, we have
\begin{equation}
\chi_{EB} = S(AB) - S(A|B).
\end{equation}
$S(AB)$ and $S(A|B)$ can be calculated from the symplectic eigenvalues $\nu_k$ of the respective covariance matrix, $\Gamma_{AB}$ and $\Gamma_{A|b}$, via the relation $S(\cdot)=\sum_{k=1}^N \frac{\nu_k+1}{2}\log_2 \frac{\nu_k+1}{2} - \frac{\nu_k-1}{2}\log_2 \frac{\nu_k-1}{2},$ where $N$ is the number of modes~\cite{PhysRevA.59.1820}.

\section{\label{sec:higherscissors}Higher-order quantum scissors}

In this section, we show that the quantum scissors can be used to distill entanglement and generalise to higher order, focusing on the three-photon scissor from Ref.~\cite{PhysRevA.102.063715}.

First recall from Ref.~\cite{neergaard-nielsen2013quantum} that tele-amplification devices perform NLA on a restricted set of superposition sates consisting of coherent states on a ring in phase space. The devices are loss tolerant meaning that they can faithfully teleport and amplify classical mixtures of coherent states. When used in this way, the device is referred to as a loss-tolerant quantum relay.

\begin{figure}
\centering
\includegraphics[width=0.45\linewidth]{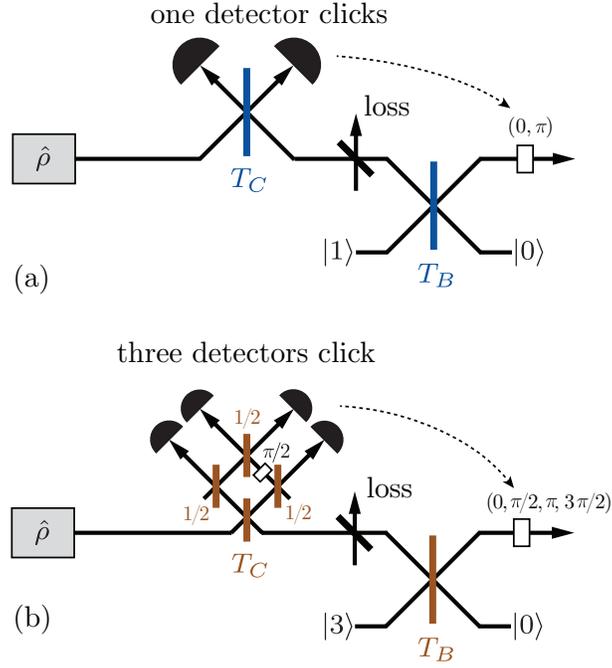}
\caption{Quantum scissors perform ideal truncation and noiseless linear amplification up to some fock number. Shown here are two examples. (a) single-photon scissor~\cite{pegg1998optical} which truncates all terms greater than Fock-number one, and (b) three-photon scissor~\cite{PhysRevA.102.063715} which truncates all terms greater than Fock-number three. A passive phase-shift (white rectangle) correction is required on the outgoing mode, depending on which detectors fire~\cite{PhysRevA.102.063715}. Note there is an important $\pi/2$ phase shift in the three scissor interferometer. An extremely useful feature is that the quantum scissors perform remarkably well for thermal input states if there is loss on the entanglement resource, as shown in the figure. We say that the scissor is loss tolerant.}
\label{fig:scissor}
\end{figure}

Quantum scissors perform simultaneous truncation and NLA up to fock number $k$ (called a $k$-photon scissor or a $k$-scissor). The first and third-order quantum scissors are shown in~\cref{fig:scissor}. Tele-amplification devices are exactly quantum scissors as the cat amplitude goes to zero. Therefore, quantum scissors are approximately loss tolerant (strictly loss tolerant only on the vacuum state). Thus, quantum scissors can be used as loss-tolerant quantum repeaters for CV QKD. We call them repeaters rather than relays since they repeat effectively and can overcome the repeaterless bound. If the input states are restricted to classical mixtures of coherent states, the device performs pretty-good NLA despite loss on the entanglement mode, as shown in~\cref{fig:scissor}.

We want to be clear that scissors perform ideal NLA up to some Fock number but the NLA is not perfect if the scissor is used as a quantum repeater. It is always better in terms of fidelity to use a lossless quantum scissor (i.e., moving the repeater all the way into Bob's station), however, by sacrificing some fidelity by using the scissor as a repeater improves the scaling of the probability of success and increases the secret key rate.

The entanglement of formation quantifies the minimum entanglement needed to prepare an entangled state from a classical one. We calculate the Gaussian entanglement of formation~\cite{PhysRevA.69.052320} as an entanglement measure using results from Ref.~\cite{PhysRevA.96.062338}. The results are presented in the main text.

\begin{figure}
\includegraphics[width=0.55\linewidth]{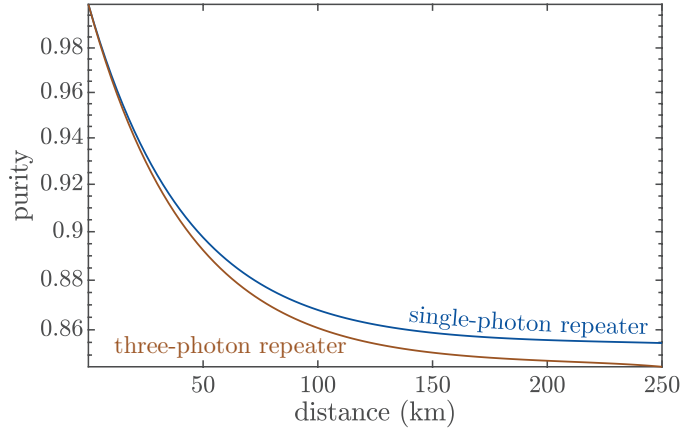}
\caption{Purity versus distance for our repeater protocol shown in~\cref{fig:protocol}(b) of the main text in symmetric configuration ($T_C{=}T_B{=}0.5,\eta_A{=}\eta_B{=}\sqrt{\eta}$), two-mode-squeezing parameter $\chi=0.2$, and with excess noise. The purity for infinite two-mode-squeezing is not shown because it decreases exponentially with distance whereas, with the repeater, the purity stays high with distance.}
\label{fig:purity_3scissor}
\end{figure}

Another important figure of merit is the purity. The purity of $\hat{\rho}_{AB}$ is defined as $\Tr({\hat{\rho}_{AB}^2})$. In~\cref{fig:purity_3scissor}, we plot the purity calculated from the density matrix, $\hat{\rho}_{AB}$, as a function of distance.

\section{\label{sec:three-repeater}Three-repeater chain}

In this section, we describe how to calculate secret key rates of the three-repeater chain shown in~\cref{fig:rate_4links} of the main text.

We use quantum memories to store the heralded state of our single-node protocol, then do entanglement swapping on two copies of the state using another quantum scissor. We calculate the output state $\hat{\rho}_{AB}$ numerically in Fock space.

The rate, $R$, of successful operation of the entire three-repeater protocol depends on the success probabilities of the higher and lower levels in the following way:
\begin{align}
R &= \frac{1}{Z_1(P_\text{lower level})} P_\text{higher level},
\end{align}
where $P_\text{higher level}$ is the success probability of the entanglement-swapping scissor (red), and $P_\text{lower level}$ is the minimum success probability of the two entanglement-distillation scissors (blue). The function $Z_n(P)$ is the average number of steps required to generate successful outcomes in $2^n$ probabilistic operations, each with success probability $P$~\cite{PhysRevA.83.012323}:
\begin{align}
    Z_n(P) &= \sum_{j=1}^{2^n} {n \choose j} \frac{(-1)^{j+1}}{1-(1-P)^j}.
\end{align}

For the same number of resources, our protocol scales like the square root of the scaling of previous schemes~\cite{PhysRevResearch.2.013310,Dias_2020,Ghalaii_2020} (we can go twice as far with the same number of resources). That is, the scheme from~\cite{Dias_2020} requires four memories, two scissors, and a Bell measurement to scale like the single-repeater bound and to beat the PLOB bound at about 400 km, whereas, our protocol uses the same resources to achieve three-repeater scaling and we beat the bound at 234 km and is robust to some excess noise and experimental imperfections.

\subsection{Asymmetric configuration}

In~\cref{fig:rate_4links} of the main text, we tune the beamsplitters and $\chi$ so that the key is roughly optimised. That is, the two-mode-squeezing parameter is $\chi=0.2$, which is a bit too large for the protocol so the effective change in amplitude of the total channel is $0.42$, set by the location of the repeaters and the transmissivity of the primary gain beamsplitter (with transmissivity $0.73$, marked with a $*$ in~\cref{fig:repeater} of the main text), and all other beamsplitters are $50{:}50$, that is, the second two links are in symmetric configuration while the first two links are in asymmetric configuration, and the central node is positioned in the exact centre between Alice and Bob.



%

\end{document}